\documentclass[twocolumn,english,aps,prb,twocolum,superscriptaddress,bibnotes,amsmath,amssymb,floatfix]{revtex4-1}
\usepackage[colorlinks=true,citecolor=blue,linkcolor=magenta]{hyperref}

\usepackage[markup=nocolor, authormarkupposition=left]{changes} 
\usepackage{soul}
\usepackage[utf8]{inputenc}
\usepackage[english]{babel}
\usepackage{amsmath,amsfonts,amssymb}
\usepackage[T1]{fontenc}
\usepackage{url}
\usepackage{changes}

\usepackage{amsmath}
\usepackage{amsfonts}
\usepackage{amssymb}
\usepackage{siunitx}
\usepackage{epstopdf}
\usepackage{graphicx}
\graphicspath{{./Figures/}}

\usepackage{changes}
\usepackage{soul}

\begin{document}

\title{Low-loss integrated nanophotonic circuits with layered semiconductor materials}

\author{Tianyi Liu}
\thanks{These authors contributed equally to this work.}
\affiliation{Institute of Physics, Swiss Federal Institute of Technology Lausanne (EPFL), CH-1015 Lausanne, Switzerland}

\author{Ioannis Paradisanos}
\thanks{These authors contributed equally to this work.}
\affiliation{Cambridge Graphene Centre, University of Cambridge, Cambridge CB3 0FA, UK}

\author{Jijun He}
\affiliation{Institute of Physics, Swiss Federal Institute of Technology Lausanne (EPFL), CH-1015 Lausanne, Switzerland}

\author{Alisson R. Cadore}
\affiliation{Cambridge Graphene Centre, University of Cambridge, Cambridge CB3 0FA, UK}

\author{Junqiu Liu}
\affiliation{Institute of Physics, Swiss Federal Institute of Technology Lausanne (EPFL), CH-1015 Lausanne, Switzerland}

\author{Mikhail Churaev}
\affiliation{Institute of Physics, Swiss Federal Institute of Technology Lausanne (EPFL), CH-1015 Lausanne, Switzerland}

\author{Rui Ning Wang}
\affiliation{Institute of Physics, Swiss Federal Institute of Technology Lausanne (EPFL), CH-1015 Lausanne, Switzerland}

\author{Arslan S. Raja}
\affiliation{Institute of Physics, Swiss Federal Institute of Technology Lausanne (EPFL), CH-1015 Lausanne, Switzerland}

\author{Clément Javerzac-Galy}
\affiliation{Institute of Physics, Swiss Federal Institute of Technology Lausanne (EPFL), CH-1015 Lausanne, Switzerland}

\author{Philippe Rölli}
\affiliation{Institute of Physics, Swiss Federal Institute of Technology Lausanne (EPFL), CH-1015 Lausanne, Switzerland}

\author{Domenico De Fazio}
\affiliation{Cambridge Graphene Centre, University of Cambridge, Cambridge CB3 0FA, UK}

\author{Barbara L. T. Rosa}
\affiliation{Cambridge Graphene Centre, University of Cambridge, Cambridge CB3 0FA, UK}

\author{Sefaattin Tongay}
\affiliation{School for Engineering of Matter, Transport and Energy, Arizona State University, Tempe, AZ 85287, USA}

\author{Giancarlo Soavi}
\affiliation{Cambridge Graphene Centre, University of Cambridge, Cambridge CB3 0FA, UK}
\affiliation{Institute for Solid State Physics, Friedrich-Schiller University Jena, 07743 Jena, Germany}

\author{Andrea C. Ferrari}
\email[]{acf26@eng.cam.ac.uk}
\affiliation{Cambridge Graphene Centre, University of Cambridge, Cambridge CB3 0FA, UK}

\author{Tobias J. Kippenberg}
\email[]{tobias.kippenberg@epfl.ch}
\affiliation{Institute of Physics, Swiss Federal Institute of Technology Lausanne (EPFL), CH-1015 Lausanne, Switzerland}

\begin{abstract}
Monolayer transition metal dichalcogenides with direct bandgaps are emerging candidates for microelectronics, nano-photonics, and optoelectronics. 
Transferred onto photonic integrated circuits (PICs), these semiconductor materials have enabled new classes of light-emitting diodes, modulators and photodetectors, that could be amenable to wafer-scale manufacturing. %could be fabricated using CMOS technology with high yields and large volumes. 
For integrated photonic devices, the optical losses of the PICs are critical. 
In contrast to silicon, silicon nitride ($\mathrm{Si_3N_4}$) has emerged as a low-loss integrated platform with a wide transparency window from ultraviolet to mid-infrared and absence of two-photon absorption at telecommunication bands. 
Moreover, it is suitable for nonlinear integrated photonics due to its high Kerr nonlinearity and high-power handing capability. 
These features of $\mathrm{Si_3N_4}$ are intrinsically beneficial for nanophotonics and optoelectronics applications. 
Here we report a low-loss integrated platform incorporating monolayer molybdenum ditelluride (1L-MoTe$_2$) with $\mathrm{Si_3N_4}$ photonic microresonators. 
We show that, with the 1L-MoTe$_2$, microresonator quality factors exceeding $3\times10^{6}$ in the telecommunication  O-band to E-band are maintained. 
We further investigate the change of microresonator dispersion and resonance shift due to the presence of 1L-MoTe$_2$, and extrapolate the optical loss introduced by 1L-MoTe$_2$ in the telecommunication bands, out of the excitonic transition region. 
Our work presents a key step for low-loss, hybrid PICs with layered semiconductors without using heterogeneous wafer bonding.
\end{abstract}
\maketitle

%%%%%%%%%%%%%%%%
%Here we mainly limit our study in the telecom wavelength, which the Si photonics is operated (between 1300 to 1600)

Layered materials (LMs)\cite{Ferrari:15},  such as graphene, transition metal dichalcogenides (TMDs), and black phosphorus, are emerging platforms for applications in microelectronics\cite{Wang:12, Manzeli:17, Hui:17}, nanophotonics\cite{Xia:14, Krasnok:18, Romagnoli:18, Bonaccorso:10}, and optoelectronics\cite{Mueller:10, Sun:16, Koppens:14, Mak:16}. 
Particularly important for optoelectronics are LMs with direct bandgaps, such as atomically thin, monolayer (1L) TMDs, that have opened up new physics and applications such as valleytronics\cite{Schaibley:16}, spintronics\cite{Zibouche:14}, transistors\cite{Radisavljevic:11}, light-emitting diodes (LEDs)\cite{Withers:15}, modulators\cite{Sun:16}, and photodetectors \cite{Koppens:14}. 
Semiconducting TMDs undergo an indirect-to-direct bandgap transition when exfoliated from bulk to monolayer\cite{Novoselov:05, Mak:10}, and their optical properties are dominated by excitonic transitions\cite{WangRMP:18, Barbone:18, Paradisanos:20}, typically in the visible or near-infrared wavelengths . 
For example, monolayer molybdenum diselenide (1L-MoSe$_2$) has a bandgap of $\sim$1.7 eV\cite{Ross:13} ($\sim$730 nm), and monolayer molybdenum ditelluride (1L-MoTe$_2$) has a bandgap of $\sim$1.1 eV\cite{Ruppert2014, Robert2016} ($\sim$1130 nm). 
%This family of materials can be integrated in optical cavities to enhance light-matter interactions and many-body effects\cite{Koppens:14, Mak:16,Engel:12, Ross:14, Bie:17}.
%MuenchNanoLett2019

Strong light-matter interactions with LMs is beneficial to optoelectronics applications. 
One approach to enhance light-matter interactions is to integrate LMs with low-loss photonic microresonators\cite{Engel:12, Ross:14, Bie:17, Javerzac-Galy:18, Datta:20}. 
For example, high-responsivity photodetectors employ resonant structures to significantly enhance the photon absorption. %modulator
The performance depends not only on the microresonator's mode volume, but also on the quality factor defined as $Q=\omega/\kappa$, where $\omega/2\pi$ is the frequency of the optical mode and $\kappa/2\pi$ is the resonance linewidth. 
Physically, the $Q$ factor is a dimensionless parameter describing the energy stored in the microresonator divided by the energy dissipation per round trip, thus the $Q$ factor represents the power storing capability of a microresonator (i.e. the intracavity power enhancement to the external driving)\cite{Vahala:03}. 
In addition, the loaded $Q$ is related to the intrinsic quality factor $Q_{0}$ (determined by the light absorption and scattering within the microresonator) and the external coupling $Q_\text{ex}$ (determined by how strong the microresonator is coupled to the external driving)\cite{Cai:00}, as $Q^{-1}=Q_0^{-1}+Q_\text{ex} ^{-1}$. 
%The Purcell factor $F_\text{p}$, i.e. the enhancement of spontaneous emission inside a cavity\cite{Vahala:03}, is related to $Q$ as $F_\text{p}=(3Q\lambda^{3})/(4\pi^2V_\text{eff})$, where $\lambda$ is the photon wavelength and $V_\text{eff}$ the mode volume of the confined field.

\begin{figure*}[t!]
\centering
\includegraphics[clip,scale=1]{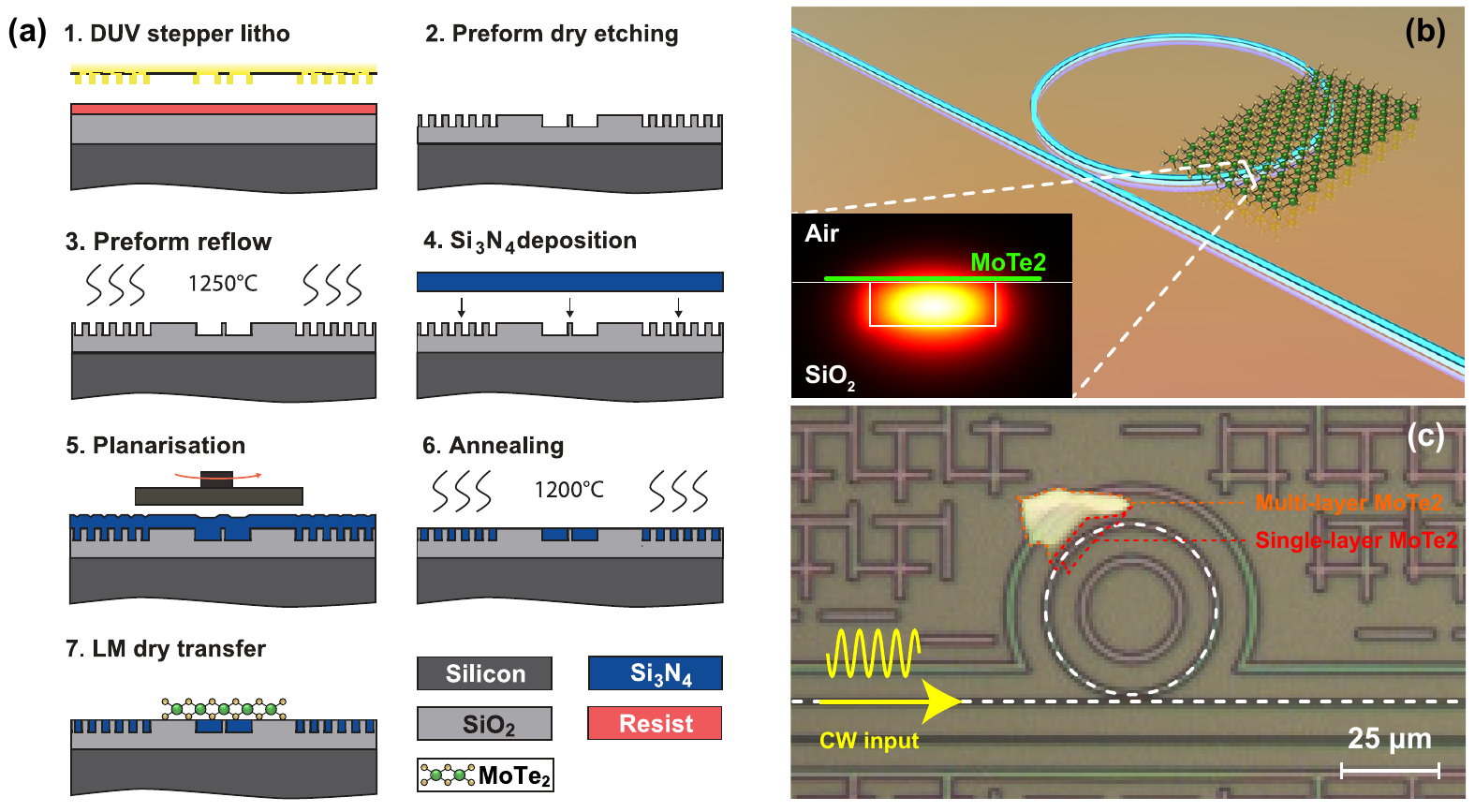}
\caption{
\textbf{Fabrication process flow and the sample overview}. 
(a) The photonic Damascene reflow process to fabricate air-cladded, high-$Q$, Si$_3$N$_4$ microresonators and the integration of 1L-MoTe$_2$. 
1. Deep-UV stepper lithography is used to pattern the Si$_3$N$_4$ structures. 
2. The pattern is transferred from the photoresist mask to the SiO$_2$ substrate via dry etching. 
3. The preform reflow at 1250$^\circ$C to reduce the waveguide sidewall roughness. 
4. LPCVD Si$_3$N$_4$ is deposited on the patterned substrate, filling the trenches and forming the Si$_3$N$_4$ structures. 
5. The excess Si$_3$N$_4$ is removed by CMP, ensuring a flat top surface. 
6. The substrate is annealed at 1200$^\circ$C to drive out the residual hydrogen content in Si$_3$N$_4$. 
7. Dry transfer of LMs onto the Si$_3$N$_4$ microresonator. 
(b) Schematic of the Si$_3$N$_4$ microring resonator covered with 1L-MoTe$_2$. 
(c) Optical microscope image showing the Si$_3$N$_4$ microring resonator, the bus waveguide, and the MoTe$_2$. 
The white dashed line marks the Si$_3$N$_4$ ring resonator and bus waveguide, where light propagates.}
\label{Fig:PF}
\end{figure*}

Planar photonic integrated circuits (PICs) are ideal platforms to enhance light-matter interactions with 1L-TMDs, and to build new classes of high-performance integrated devices such as LEDs, modulators, and photodetectors. 
The planarized top surfaces of waveguides and microresonators are advantageous, as the 1L-TMDs' flatness is maintained and the strain-induced local fluctuations in the 1L-TMDs' band structure\cite{Dhakal:17} are avoided. 
With deterministic transfer techniques of 1L-TMDs, these devices could be fabricated using CMOS technology with high yields and large volumes. 
For most 1L-TMDs, their bandgaps for excitonic transitions lie in the visible wavelengths, while silicon, commonly used in integrated photonics, has a transparency window\cite{Soref:06} above 1.2 $\mu$m. 
%($\sim$0.15 - 1.03 eV, corresponding to 1.2 - 8 $\mu$m wavelength
This difference makes silicon-based devices challenging for applications using TMDs, such as LEDs at visible wavelengths. 

In this case, Si$_3$N$_4$\cite{Moss:13, Blumenthal:20} is a promising alternative to silicon. 
Silicon nitride has a bandgap of $\sim$5 eV, enabling a transparency window covering 400 nm to 2500 nm spectral range and absence of two-photon absorption in the telecommunication bands. 
Recent advances\cite{Xuan:16, Ji:17, Ye:19b, Liu:20, Liu:20b} in fabrication have achieved integrated Si$_3$N$_4$ waveguides with tight optical confinement and exceptionally low losses of 1 dB/m, enabling microresonators with $Q$ factors above $30\times10^6$ and 1-meter-long spiral waveguides in 20 mm$^2$ footprints. 
%The optical propagation losses in state-of-the-art integrated Si$_3$N$_4$ PIC have reached the level below 1 dB/m\cite{Spencer:14, Gundavarapu:19, Liu:20b}. 
In comparison, the lowest optical losses are still above 35 dB/m for both the silicon\cite{Selvaraja:14} and the indium phosphide\cite{Ciminelli:13}, due to the fabrication processes and the high refractive indices that enhance light scattering at the waveguide surface. 
In addition, Si$_3$N$_4$ has an excellent handling capability of high optical power potentially above 100 kW \cite{Gyger:20}. 
Combining the relatively high nonlinear index $n_2=2.5\times10^{-15}$ cm$^2$W$^{-1}$ (describing the refractive index change induced by the optical intensity) and the established geometry dispersion engineering (i.e. tailoring the group-velocity dispersion via geometry variation\cite{Foster:06}), Si$_3$N$_4$ has been used for various linear and nonlinear photonics applications, particularly for microresonator-based Kerr frequency combs\cite{Kippenberg:18} and chip-based supercontinuum generation\cite{Gaeta:19} that are based on intense light-matter interactions. 
Therefore, low-loss Si$_3$N$_4$ PICs are promising to study light-matter interactions with TMDs, and have already been used to build modulators\cite{Datta:20} and photodetectors \cite{Gonzalez-Marin:19, Gao:19}. 
These active devices are also important for Si$_3$N$_4$ photonics, as Si$_3$N$_4$ is insulating and passive, i.e. its optical properties cannot be altered via applying external electric driving to itself. 

%Active tuning of Si$_3$N$_4$ is often realized by exploiting the thermo-optic effect, which however has a low speed speed of few kilohertz \cite{Xue:16, Joshi:16}. 
%Thus, active materials, such as III-V and Si~\cite{OpdeBeeck:20, Xiang:20, Park:20}, electro-optic materials (e.g. lithium niobate~\cite{WangC:18, Wang:19}), and piezoelectric materials (e.g. lead-zirconate-titanate (PZT)~\cite{Hosseini:15, Alexander:18, Jin:18} and aluminium nitride~\cite{Stanfield:19, Tian:20, Liu:20a}), are required to realize active optoelectronic functionalities such as lasers and high-speed modulators. 

%not only resonance structure but also travelling scheme

 %Ref.\cite{liu_high-yield_2020} fabricated meter-long, Si$_3$N$_4$ WGs with optical loss down to$\sim$1dB/m\cite{liu_high-yield_2020}. 
%Such Si$_3$N$_4$ WGs could feature a longer region coupled with 1L-TMDs, while maintaining low losses, useful for (spectrally) broadband applications, such as traveling wave amplifiers\cite{GuoNatPhot2018}.

Here, we integrate Si$_3$N$_4$ microresonators with monolayer molybdenum ditelluride (1L-MoTe$_2$), and investigate the optical response of this Si$_3$N$_4$-TMD hybrid system. 
The 1L-MoTe$_2$ has an bandgap of $\sim$1.1 eV ($\sim$1130 nm\cite{Ruppert2014, Robert2016}), thus it is essentially transparent in the telecommunication bands from 1280 nm to 1630 nm. 
We use photonic chip-based Si$_3$N$_4$ microring resonators with free spectral ranges (FSR, the frequency spacing of the optical resonances) of 150 GHz and 1 THz. 
The FSR values, $D_1/2\pi$, are calculated from the resonant condition $D_1/2\pi=c/(2\pi R\cdot n_g)$, where $c$ is the speed of light in vacuum and $n_g$ is the group index of the microresonator ($c/n_g$ being the group velocity of light). 
The Si$_3$N$_4$ microresonators, fabricated using the photonic Damascene process\cite{Pfeiffer:18b}, have no top SiO$_2$ cladding, thus the 1L-MoTe$_2$ transferred onto the Si$_3$N$_4$ waveguide's top surface can directly interact with the optical mode in the waveguide core. 
%To the best of our knowledge, this $Q_{0}$ value is at least one order of magnitude higher than all previous reported LM integrated Si$_3$N$_4$ devices\cite{DattaNatPhot2020,YaoNature2018}. Our high-$Q_{0}$ devices could be exploited for generation of soliton microcombs using compact III-V lasers\cite{SternNature}, while allowing on-chip in situ dispersion control. 

\begin{figure*}[t!]
\centering
\includegraphics[clip,scale=1]{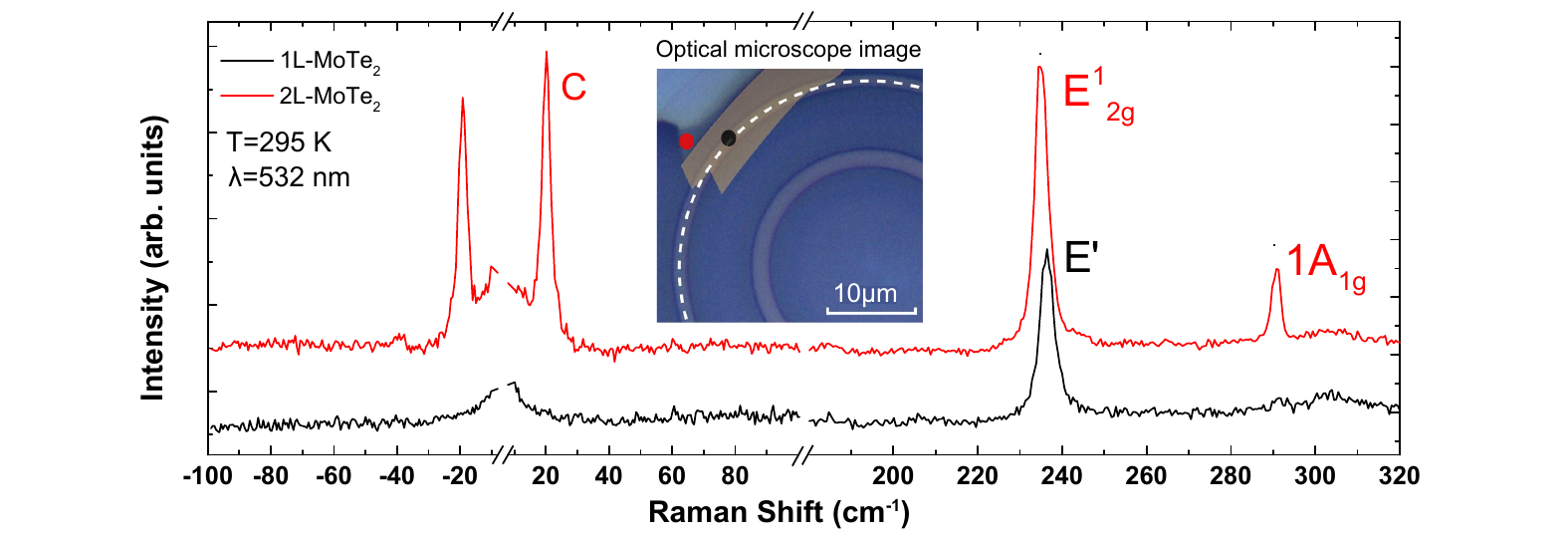}
\caption{
\textbf{Raman spectra of the 1L-MoTe$_2$ (black) on microresonator and the 2L-MoTe$_2$ (red) outside it}. 
Inset: optical microscope image of the 1L-MoTe$_2$ on the Si$_3$N$_4$ microresonator. 
The black and red spots indicate the collection areas of the representative Raman spectra of 1L- and 2L-MoTe$_2$. 
The white dashed line marks the Si$_3$N$_4$ microresonator, and the grey area is the 1L-MoTe$_2$ flake. 
%IMC: in-plane metal-chalcogen vibrational mode. 
%OMC: out-of-plane metal-chalcogen vibrational mode.
}
\label{Fig:Raman}
\end{figure*}

Figure \ref{Fig:PF}(a) presents the fabrication process flow of our Si$_3$N$_4$ PIC with LM. 
The Si$_3$N$_4$ PIC is fabricated using the photonic Damascene process\cite{Pfeiffer:18b}. 
Deep-UV stepper lithography based on KrF at 248 nm is used to pattern Si$_3$N$_4$ waveguides, microresonators, and stress-release structures\cite{Pfeiffer:18b} (to prevent crack formation in Si$_3$N$_4$ due to the tensile film stress). 
The pattern is transferred from the photoresist mask to the SiO$_2$ substrate using dry etching based on C$_4$F$_8$, O$_2$ and He, followed by a preform reflow\cite{Pfeiffer:18} at 1250$^{\circ}$C, to reduce the surface roughness of the waveguide. 
Low-pressure chemical vapor deposition (LPCVD) based on the precursors SiH$_2$Cl$_2$ and NH$_3$ is used to deposit high-quality Si$_3$N$_4$ on the patterned preform, filling the trenches and forming the waveguide cores. 
The excess Si$_3$N$_4$ is removed by chemical-mechanical polishing (CMP), creating a flat top surface with a root-mean-square (RMS) roughness below 0.3 nm \cite{Pfeiffer:18}, ideal to maintain the 1L-TMDs' flatness. 
The substrate is then annealed at 1200$^{\circ}$C to drive out the residual hydrogen introduced during the LPCVD process. 
We note a height difference of less than 40 nm between the Si$_3$N$_4$ waveguide cores and their surrounding SiO$_2$ cladding, measured by scanning electron microscopy (SEM) and atomic force microscopy (AFM). 
This height difference is caused by the post-CMP cleaning using buffered HF to removed the CMP slurry particles, and can be reduced in the future by optimizing the cleaning process. 
Nevertheless, as illustrated below, the Si$_3$N$_4$ waveguides without top SiO$_2$ cladding allows seamless contact of 1L-MoTe$_2$ with the optical mode via evanescent coupling. 
This feature is a key advantage of the photonic Damascene process due to its additive fabrication nature. 
In comparison, the top-down, subtractive process\cite{YangY:18, Wu:20} requires complex control of the planarization, in order to create a flat wafer surface with bare waveguides. 

Monolayer MoTe$_2$ flakes are exfoliated from bulk 2H-MoTe$_2$ crystals (alpha phase) prepared by flux zone growth\cite{Zhang:15} on Nitto Denko tape\cite{Novoselov:05}, and then exfoliated again on a polydimethylsiloxane (PDMS) stamp placed on a glass slide for inspection under an optical microscope. 
Optical contrast is optimized to identify the monolayer prior to transfer\cite{Casiraghi:07}. 
Before 1L-MoTe$_2$ transfer, the Si$_3$N$_4$ microresonators are wet-cleaned by 60 s ultrasonication in acetone and isopropanol, and exposed to oxygen-assisted plasma at 10W for 60s. 
The 1L-MoTe$_2$ flakes are aligned and stamped on the Si$_3$N$_4$ microresonators with a micro-manipulator at 40$^{\circ}$C, before increasing the temperature to 60$^{\circ}$C, so the flakes detach from the PDMS and adhere preferentially to the microresonators\cite{Orchin:19}. 
Figure \ref{Fig:PF}(b) illustrates a Si$_3$N$_4$ microring resonator covered with 1L-MoTe$_2$. 
Figure \ref{Fig:PF}(c) shows the optical microscope image of the sample, with the white dashed line marking the Si$_3$N$_4$ bus waveguide and the microring resonator. 
Light is coupled from the bus waveguide to the microring resonator via in-plane evanescent coupling with a gap distance of 775 nm.

To characterize the transferred 1L-MoTe$_2$, we use micro-Raman spectroscopy at 532 nm ($\sim$2.3 eV), close to the C-exciton energy of MoTe$_2$\cite{Broomley1972bands}, to enhance the electron-phonon interaction\cite{Song:16, Trovatello:20} in the vicinity of the $\Gamma$-point of the Brillouin zone. 
Less than 50 $\mu$W power is used to avoid thermal effects. 
The measurements are performed in a Horiba LabRam Evolution with a cut-off frequency of $\sim$5 cm$^{-1}$, a 1800 l/mm grating and a spot size of $\sim$700 nm. 
We measure 1L-MoTe$_2$ located on the microresonator (black dot in the Fig. \ref{Fig:Raman} inset), and compare it with a 2L-MoTe$_2$ flake out of the microresonator (red dot in the Fig. \ref{Fig:Raman} inset). 
The Raman spectra of 1L- and 2L-MoTe$_2$ are also shown in Fig. \ref{Fig:Raman}. 
A strong C mode\cite{Tan:12, Zhang:13} is observed at 19.5 cm$^{-1}$ in 2L-MoTe$_2$, but is absent for 1L-MoTe$_2$, as expected\cite{Zhang:13}. 
The in-plane modes are 236.4 cm$^{-1}$ for 1L-MoTe$_2$ (E' mode) and 234.9 cm$^{-1}$ for 2L-MoTe$_2$ (E$ _{2g}^{1}$ mode), in good agreement with literature\cite{Goldstein2016}. 
The out-of-plane metal-chalcogen vibration (B$_{2g}$ in bulk and 1A$_{1g}$ in 2L-MoTe$_2$\cite{Goldstein2016}) is only observed in 2L-MoTe$_2$ 290.8 cm$^{-1}$, because B$_{2g}$ symmetry modes are inactive for 1L-MoTe$_2$\cite{Goldstein2016}. 
This confirms the transfer of 1L-MoTe$_2$ on the microresonator.

\begin{figure*}[t!]
\centering
\includegraphics[clip,scale=1]{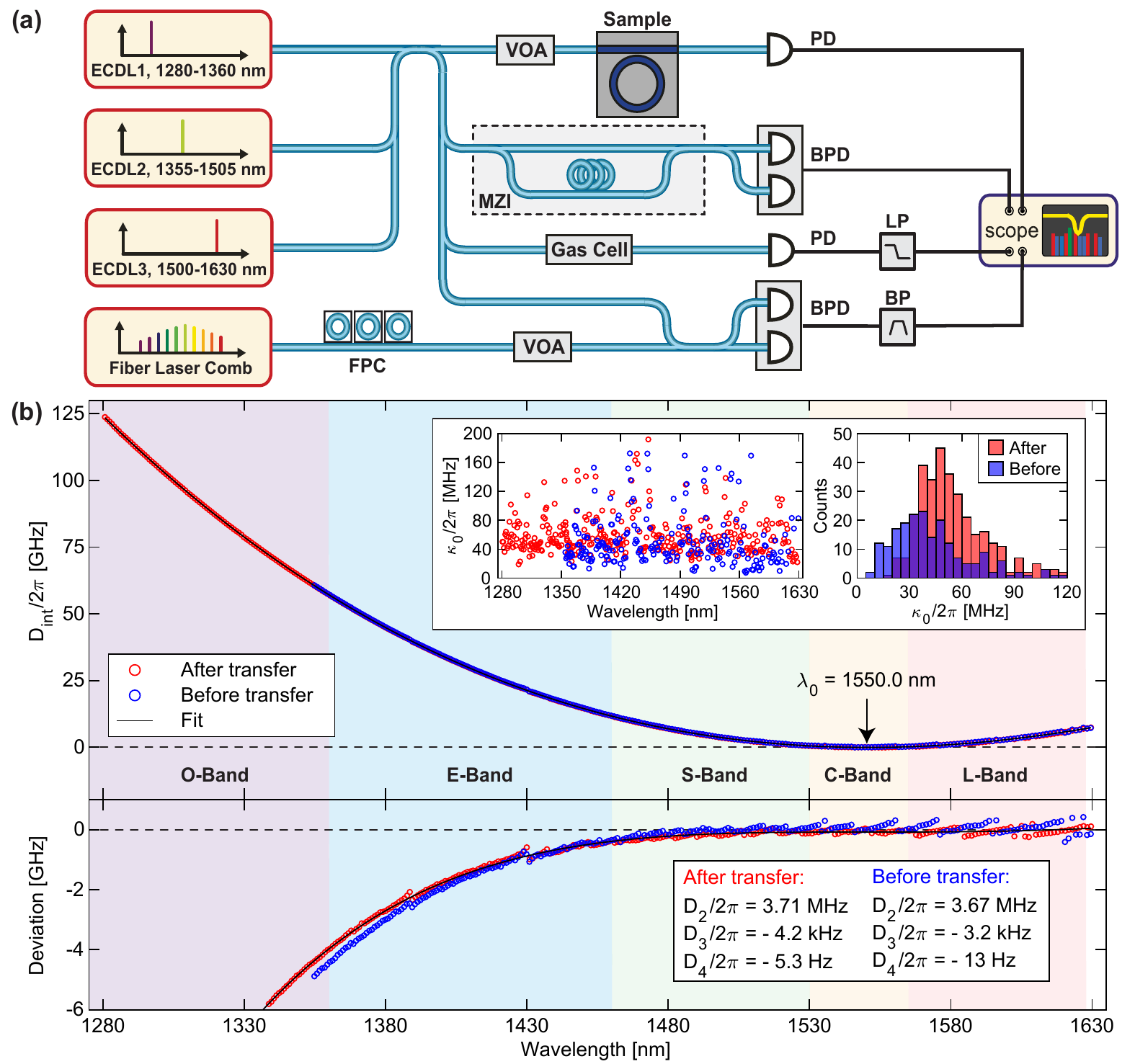}
\caption{
\textbf{Experimental characterization of the loss and dispersion of the 150-GHz-FSR Si$_3$N$_4$ microresonator}. 
(a) Experimental setup of the frequency-comb-assisted cascaded diode laser spectroscopy used to characterize the microresonator's loss and dispersion. 
ECDL: external-cavity diode laser. 
FPC: fiber polarization controller. 
VOA: variable optical attenuator. 
MZI: Mach–Zehnder interferometer. 
PD(BPD): (balanced) photodetector. 
LP/BP: electronic low/band pass filter. 
(b) Top: Measured integrated microresonator dispersion profile $D_{\mathrm{int}}/2\pi$, covering from 1280 nm to 1630 nm for the TE$_{00}$ mode. 
The reference resonance wavelength chosen here is 1550 nm. 
Insets: the $\kappa_0/2\pi$ of each resonance, and the $\kappa_0/2\pi$ histograms showing that the most probable value increases from 35 MHz to 50 MHz after the 1L-MoTe$_2$ transfer. 
Bottom: Resonance frequency deviation from the $D_2$-dominant parabolic profile $[D_\text{int}(\mu)-D_2\mu^2/2]/2\pi$, to reveal the change in higher-order dispersion (such as $D_3/2\pi$) and avoided mode crossings.}
\label{Fig:150G}
\end{figure*}

To investigate the effect of the transferred 1L-MoTe$_2$ on the Si$_3$N$_4$ optical mode, we characterize the optical resonances of the integrated hybrid microresonator system. 
Figure \ref{Fig:150G}(a) shows the experimental setup. 
Light is coupled into and out of the Si$_3$N$_4$ chip using lensed fibers and double-inverse nanotapers at the chip facets\cite{Liu:18}. 
The total fiber-chip-fiber through coupling efficiency, i.e. the ratio of the optical power in the output fiber to the power in the input fiber, is approximately 40$\%$. 
To measure the resonance frequencies and linewidths (i.e. loss of each resonance), the optical transmission spectrum of the microresonator is calibrated using a frequency-comb-assisted diode laser spectroscopy\cite{DelHaye:09, Liu:16}. 
Three mode-hop-free, widely tunable, external-cavity diode lasers (ECDLs) are used and cascaded\cite{Liu:16}, to cover the telecommunication O-band to E-band (1280 nm to 1630 nm). 
When scanning the laser frequency, the instantaneous laser frequency is acquired by beating the laser with a commercial, self-referenced, fiber-laser-based optical frequency comb\cite{DelHaye:09}, assisted further with a Mach-Zehnder interferometer\cite{Pfeiffer:18b}. 
In addition, molecular absorption spectroscopy is performed during the laser frequency scan\cite{Liu:16}, which is used to further calibrate the optical transmission spectrum of the microresonator to precisely extract the absolute frequency of each recorded data point. 
%is done using both the beating signal, with a commercial fiber-laser frequency comb (MenloSystems), and a Mach-Zehnder interferometer signal. 
%A gas cell is used to acquire the frequency of the whole transmission trace, by comparing the absorption during scanning with the high-resolution transmission molecular absorption database\cite{hitran}. 
The light polarization is constant during the laser scan as polarization-maintaining fiber components are used. 
Optical resonances in the microresonator transmission spectrum are identified and fitted\cite{Liu:18a}, to extract the microresonator's intrinsic loss $\kappa_0/2\pi$ and the bus-waveguide-to-microresonator external coupling strength $\kappa_\text{ex}/2\pi$, respectively. 
The loaded linewidth is thus calculated as $\kappa/2\pi=\kappa_\text{ex}/2\pi+\kappa_0/2\pi$ ($\kappa$, $\kappa_\text{ex}$ and $\kappa_0$ are angular frequencies).
%this gives the linewidth\cite{Liu2016,Liu2018b}.

\begin{figure*}[t!]
\centering
\includegraphics[clip,scale=1]{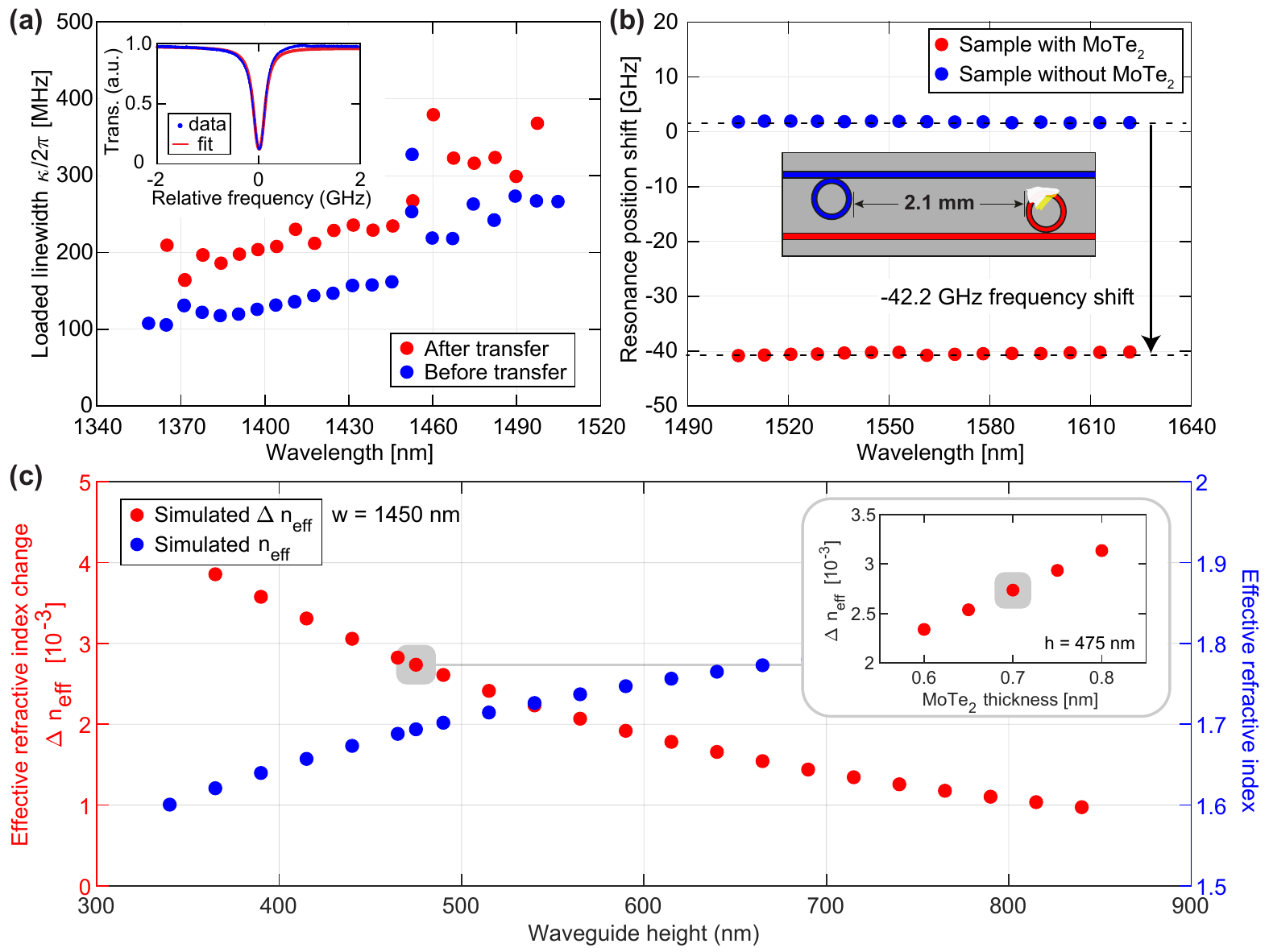}
\caption{
\textbf{Experimental characterization and simulations of the loss and resonance frequency shift of the 1-THz-FSR Si$_3$N$_4$ microresonator with 1L-MoTe$_2$}.
(a) The measured loaded resonance linewidth $\kappa/2\pi$ before and after the 1L-MoTe$_2$ transfer. 
Inset: a representative resonance profile at 1425 nm and its Lorentzian fit, giving $\kappa_{\mathrm{ex}}/2\pi=82$ MHz and $\kappa_0/2\pi=147$ MHz. 
(b) The measured and calculated absolute frequency shift of the TE$_{00}$ resonance grid before and after the 1L-MoTe$_2$ transfer, for two microresonators, one with 1L-MoTe$_2$ and the other without. 
Inset: schematic showing that the two microresonators are distant by 2.1 mm on the same chip. 
(c) Simulated effective refractive index $n_{\mathrm{eff}}$ and the change $\Delta n_{\mathrm{eff}}$ of the effective refractive index due to the presence of 1L-MoTe$_2$, as a function of the Si$_3$N$_4$ waveguide height. 
The simulated wavelength is 1550 nm in the TE$_{00}$ mode in a 1-THz-FSR microresonator. 
Insets: simulated $\Delta n_{\mathrm{eff}}$ as a function of the 1L-MoTe$_2$ thickness.}
\label{Fig:1T}
\end{figure*}

Monolayer MoTe$_2$ is transferred onto Si$_3$N$_4$ microresonators with two FSRs, 150 GHz and 1 THz. 
First, we characterize the 150-GHz-FSR microresonator. 
Here we mainly study the transverse-electric fundamental mode of the microresonator, i.e. the TE$_{00}$ mode. 
We note that, the fundamental mode in the other polarization -- the transverse-magnetic (TM$_{00}$) -- typically has higher losses than the TE$_{00}$ mode.
We characterize the same Si$_3$N$_4$ microresonator before and after the 1L-MoTe$_2$ transfer, for comparison and to extract the extra loss introduced by the 1L-MoTe$_2$ transfer. 
The measured intrinsic losses $\kappa_0/2\pi$ are plotted in the Fig. \ref{Fig:150G}(b) inset. 
No wavelength-dependent $\kappa_0/2\pi$, before and after the transfer, is observed far above the 1L-MoTe$_2$'s excitonic resonance at $\sim$1130 nm\cite{Ruppert2014}. 
Histograms of $\kappa_0/2\pi$ show that the most probable value increases from 35 MHz to 50 MHz after the transfer. 
This loss increment corresponds to a $Q_0$ degradation from $5.5\times10^6$ to $3.9\times10^6$, representing a viable low-loss microresonator with TMD integration. 

The measured integrated microresonator dispersion, defined as $D_{\mathrm{int}}(\mu)=\omega_\mu-\omega_0-\mu D_1=D_2\mu^2/2+D_3\mu^3/6+D_4\mu^4/24+...$, is shown in Fig. \ref{Fig:150G}(b) top. 
Here $\omega_\mu/2\pi$ is the frequency of the $\mu^{\mathrm{th}}$ resonance, $D_1/2\pi$ is the microresonator FSR, $D_2/2\pi$ is the group-velocity dispersion, $D_3/2\pi$ and $D_4/2\pi$ are higher-order dispersion terms. 
The reference resonance frequency is $\omega_0/2\pi=193.4$ THz (corresponding to $\lambda_0=1550$ nm). 
Limited by the size of the transferred flake, the $D_{\mathrm{int}}$ curves before and after the transfer are nearly identical. 
To further investigate the dispersion change, the deviation from the $D_2$-dominant parabolic profile, defined as $[D_{\mathrm{int}}(\mu)-D_2\mu^2/2]/2\pi$, is shown in Fig. \ref{Fig:150G}(b) bottom. 
Higher-order dispersion change (e.g. $D_3/2\pi$ and $D_4/2\pi$) and avoided mode crossings\cite{Herr2014} are revealed. 
The measured parameters from fitting are, $D_2/2\pi=3.67$ MHz and $D_3/2\pi=-3.2$ kHz before the 1L-MoTe$_2$ transfer, and $D_2/2\pi=3.71$ MHz and $D_3/2\pi=-4.2$ kHz after the 1L-MoTe$_2$ transfer.

To confirm the coupling of the Si$_3$N$_4$ optical mode to the 1L-MoTe$_2$, the same device is  further characterized in the wavelength range of 1050 nm to 1090 nm (near the 1L-MoTe$_2$ excitonic resonance at $\sim$1130 nm),  using another ECDL operating in this wavelength range. 
A different microresonator with a gap distance of 370 nm between the bus waveguide and the microresonator is used, in which the small gap distance ensures strong evanescent coupling between the the bus waveguide and the microresonator in this wavelength range. 
Before the 1L-MoTe$_2$ transfer, the mean loaded resonance linewidth is $\kappa/2\pi\approx200$ MHz. 
After the transfer, most of the resonances, which were previously visible, vanish. 
The remaining resonances detected have $\kappa/2\pi>1$ GHz, corresponding to an \textit{average} loss of 1.5 dB/cm near 1 $\mu$m wavelength . 
Note that, this value is averaged over the entire microresonator, despite the fact that only a small part of the microresonator is covered by 1L-MoTe$_2$. 
Comparing with the optical loss measured at the telecommunication bands of 1280 nm to 1630 nm, this wavelength-dependent loss is consistent with the significant optical absorption of $\sim3.4\%$ at the 1L-MoTe$_2$ band edge\cite{Ruppert2014}, and confirms that the 1L-MoTe$_2$ is coupled to the optical mode. 

Next, we characterize the 1-THz-FSR microresonator (with a ring radius of 22 $\mu$m). 
The smaller microresonator size leads to a larger ratio of its perimeter covered by the 1L-MoTe$_2$. 
Here we estimate the ratio as $\eta\approx13$\%. 
Figure \ref{Fig:1T}(a) shows the loaded linewidths $\kappa/2\pi$ of the measured TE$_{00}$ resonances from 1355 nm to 1505 nm, before and after the 1L-MoTe$_2$ transfer. 
The observed wavelength-dependence is due to the bus-waveguide-to-microresonator evanescent coupling (i.e. $\kappa_\text{ex}/2\pi$ is wavelength-dependent, and is stronger at longer wavelengths). 
For each resonance, due to the 1L-MoTe$_2$ transfer, a linewidth increment $\Delta\kappa/2\pi$ of up to 80 MHz is observed. 
As the 1L-MoTe$_2$ is atomically thin (less than 1 nm), and should not effect the bus-waveguide-to-microresonator evanescent coupling, we attribute this $\kappa/2\pi$ increment as a result of the $\kappa_0/2\pi$ increment. 
Figure \ref{Fig:1T}(a) inset shows a resonance of 1425 nm wavelength, after the 1L-MoTe$_2$ transfer, where the Lorentzian fit gives $\kappa_\text{ex}/2\pi=82$ MHz and $\kappa_0/2\pi=147$ MHz, respectively. 
This observation indicates that the 1L-MoTe$_2$ transfer introduces an extra loss to the microresonator. 
The estimated microresonator $Q$ factor is $Q_{0}=1.3\times10^6$, dropped from the initial value of $Q_{0}=3.8\times10^6$. 
The \textit{average} loss increment is estimated as 0.15 dB/cm.    
Therefore, the estimated linear optical loss in the waveguide fully covered by 1L-MoTe$_2$ is $\alpha\approx0.15\times(13\%)^{-1}$ dB/cm $=1.2$ dB/cm. 

We also investigate the frequency shift of the resonance grid induced by the 1L-MoTe$_2$. 
As mentioned earlier, the absolute frequency of each resonance is calibrated using molecular absorption spectroscopy. 
Another identical microresonator is used here, which is on the same chip but 2.1 mm distant from the one with 1L-MoTe$_2$. 
The absolute frequency of each TE$_{00}$ resonance is measured, before and after the 1L-MoTe$_2$ transfer, for both microresonators. 
The relative frequency shift of each resonance, before and after the transfer, is plotted in Fig. \ref{Fig:1T}(b), for each microresonator. 
The microresonator without 1L-MoTe$_2$ serves as a reference, and only a small frequency shift of 1.8 GHz is observed, which is likely due to the environmental temperature change when the measurements were performed. 
Note that the thermo-optic coefficient of Si$_3$N$_4$\cite{Arbabi:13} is $dn(\text{Si}_3\text{N}_4)/dT=2.5\times 10^{-5}$/K, thus the temperature-induced frequency shift coefficient is around $-2.4$ GHz/K (negative sign indicates the red shift).
%and to dust particles landed on chip before wet cleaning for 1L-MoTe$_2$ transfer. 
For the microresonator with 1L-MoTe$_2$, an average frequency shift of $-40.4$ GHz is observed. 
Including the 1.8 GHz frequency shift identified in the other microresonator, the total frequency shift due to 1L-MoTe$_2$ is $-42.2$ GHz.
Since a part of the light field is coupled into the 1L-MoTe$_2$, this $-42.2$ GHz frequency shift is induced by the increment of the effective refractive index. 
%(i.e. the phase delay per unit length in the WG, relative to that in vacuum, for the propagating mode\cite{Saleh2019}) at longer wavelengths, where the 1L-MoTe$_2$ absorption is negligible\cite{Ruppert2014}. 

Finite-element-method (FEM) simulations of the TE$_{00}$ mode using COMSOL Multiphysics are performed, to study the effective refractive index $n_\text{eff}$, and the change of the effective refractive index $\Delta n_\text{eff}$ due to the presence of 1L-MoTe$_2$. 
Figure \ref{Fig:1T}(c) shows the simulated $n_\text{eff}$ and $\Delta n_\text{eff}$ as a function of the Si$_3$N$_4$ waveguide height at 1550 nm wavelength. 
Here the waveguide has a fixed width of $w=1.45$ $\mu$m and a varying height $h$ from 340 nm to 840 nm. 
%to examine the variation in $\Delta n$ in the vicinity of the resonance characterization data. 
We include in the simulations the 40 nm height difference between the Si$_3$N$_4$ waveguide core and the SiO$_2$ cladding. 
Surface current density boundary conditions are used to model 1L-MoTe$_2$. 
We use $n(\text{MoTe}_2)=4.43$\cite{Ruppert2014}, and set the layer thickness to 0.7 nm (close to the experimental value from the AFM measurements\cite{Ruppert2014, Xia:14}). 

The experimentally measured, \textit{average} refractive index change is $\Delta n_\text{eff, avg}=n_\text{eff}\cdot\Delta f/f_0$, where $f_0$ is the resonance frequency and $ \Delta f$ is the frequency shift. 
Considering the fact that not the entire Si$_3$N$_4$ microresonator is fully covered by 1L-MoTe$_2$, the actual refractive index change is thus calculated as $\Delta n_\text{eff}=\Delta n_\text{eff, avg}/\eta$, where $\eta=13$\% is the ratio of the microresonator perimeter covered by 1L-MoTe$_2$. 
%\hl{To calculate the integrated $\Delta n_\text{eff}$ (considering the fact that the entire Si$_3$N$_4$ microresonator is not fully covered by 1L-MoTe$_2$), we use the measured $\Delta n_\mathrm{eff}=\frac{n_\mathrm{eff}\Delta f}{\eta f_0}$, 
%where $\eta=13$\% is the ratio of the microresonator perimeter covered by 1L-MoTe$_2$. }
The data shown in Fig. \ref{Fig:1T}(b) gives $\Delta n_\mathrm{eff}=2.73\times10^{-3}$. 
%Due to the thermo-optic effect in Si$_3$N$_4$\cite{arbabi_measurements_2013}, an uncertainty $dn/dT=1.7\times 10^{-4}$/K exists when calculating $n_\mathrm{eff}$. 
This value corresponds to the Si$_3$N$_4$ waveguide $h=475$ nm, with $\sim$0.3\% of the optical mode overlapping the 1L-MoTe$_2$. 
In comparison, the measured waveguide height from the SEM of the sample cross-section is $h=540$ nm, corresponding to $\Delta n_\mathrm{eff}=2.23\times10^{-3}$ in Fig. \ref{Fig:1T}(c).

In conclusion, we demonstrated the integration of 1L-MoTe$_2$ on Si$_3$N$_4$ photonic microresonators with $Q_0=1.3\times 10^6$, and estimated the optical loss of 1.2 dB/cm in the waveguide fully covered by 1L-MoTe$_2$ in the telecommunication bands. 
The Si$_3$N$_4$ photonic Damascene process already proves its compatibility with the dry transfer of LMs. 
In the future, wet transfer\cite{Coleman:11} and CVD growth\cite{LeeYH:12, Kim:19, WangJ:15} can also be used which allow integrating 1L-TMDs with larger surface areas and better single-layer uniformity. 
%For photonic applications near the LMs optical gap, other LMs with gaps from 0 eV (SLG) up to $\sim$2 eV (WS$_2$)\cite{WangRMP:18} offer a wide material selection. 
Combining with low-loss Si$_3$N$_4$ spiral waveguides of extended lengths up to a meter (Ref.\cite{Liu:20b}), high-responsivity, travelling-wave photodetectors could be built\cite{Wang:13}. 
In addition, the 1L-TMDs and single-layer graphene have a third-order-nonlinear susceptibility $\chi_3\sim10^{-18}$ m$^{2}$/V$^{2}$ (Ref.\cite{DaiACS2020, SayNatComm2017}) and $\chi_3\sim10^{-15}$ m$^{2}$/V$^{2}$ (Ref.\cite{Soavi2018, Soavi2019, Lafeta2017}), significantly higher than that of Si$_3$N$_4$ $\sim10^{-21}$m$^{2}$/V$^{2}$. 
Therefore, high-$Q$ Si$_3$N$_4$ microresonators integrated with LMs could offer viable systems to realize few-photon nonlinear optics, TMD-assisted four-wave mixing, and cavity-enhanced frequency conversion\cite{Majumdar:15}.
%Ref. \cite{Marpaung:19}. 
While maintaining the intrinsic low loss, our work presents a key step for adding novel active functionalities on the Si$_3$N$_4$ photonics platform\cite{Youngblood:17}.
% and Si ($\sim10^{-19}$m$^{2}$/V$^{2}$), 

\medskip
\begin{footnotesize}

\noindent \textbf{Acknowledgments}: 
We thank Maxim Karpov for the discussion. 
The Si$_3$N$_4$ microresonator samples were fabricated in the EPFL center of MicroNanoTechnology (CMi).
%The 1L-MoTe$_2$ was prepared and transferred to the Si$_3$N$_4$ samples in the Cambridge Graphene Center
\\

\noindent \textbf{Funding Information}: 
This work was supported by funding from the European Union H2020 research and innovation programme under the Graphene Flagship grant agreement No. 696656; the Quantum Flagships, ERC Grants Hetero2D, and GSYNCOR, EPSRC Grants EP/L016087/1, EP/N010345/1, EP/K017144/1, EP/K01711X/1; and by Swiss National Science Foundation under grant agreement No. 176563 (BRIDGE). \\

\textbf{Data Availability}: 
The code and data used to produce the plots are available from \texttt{Zenodo}.

\end{footnotesize}
\bibliographystyle{apsrev4-1}
\bibliography{MoTe2nanoresonators_ref}
\end{document}